\documentclass[prl,twocolumn,aps,showpacs,amsmath,amssymb]{revtex4}
\usepackage{graphicx}
\usepackage{dcolumn}
\usepackage{bm}

\begin{document}

\title{Textures and non-Abelian vortices in atomic $d$-wave paired Fermi condensates}

\author{H.~M. Adachi$^1$, Y. Tsutsumi$^1$, J.~A.~M.  Huhtam\"{a}ki$^{1,2}$, and K. Machida$^1$}

\address{$^1$Department of Physics, Okayama University, Okayama 700-8530, Japan}
\address{$^2$Department of Applied Physics, Helsinki University of Technology, P.O. Box 5100, 02015 TKK, Finland}

\begin{abstract}
We report on fundamental properties of superfluids with
$d$-wave pairing symmetry. We consider neutral  atomic Fermi gases in a harmonic trap,
the pairing being produced by a Feshbach resonance via a $d$-wave interaction channel.
A Ginzburg-Landau (GL) functional is constructed which is symmetry constrained
for five component order parameters (OP).
We find OP textures in the cyclic phase and stability conditions for a non-Abelian fractional 1/3-vortex 
under rotation. It is proposed how to
create the intriguing 1/3-vortex experimentally in atomic gases via optical means.
\end{abstract}



\maketitle

Superfluids with $s$-wave pairing symmetry have been realized by
using a Feshbach resonance of $^6$Li atom gases at $H$=822G in 2005\cite{s-wave}.
It is natural to expect that the research front of cold atom gases develops towards realizing condensates
with higher relative angular momentum of the Cooper pairs.
In fact, much effort from theoretical and experimental sides is now focused on $p$-wave pairing in
$^6$Li  where a $p$ -wave Feshbach resonance at $H$=159G is already confirmed  to 
exist\cite{p-wave}. Although $p$-wave molecules have been formed\cite{p-wave}, 
experimental evidence of  $p$-wave superfluidity in 
$^6$Li has not reported.

At this stage it might be useful to theoretically investigate $d$-wave superfluidity in neutral Fermi gases 
in order to further motivate experimental and theoretical works towards this direction.
Hulet has  performed a coupled channels calculation and found 
a Feshbach resonance for $d$-wave channel in $^6$Li\cite{hulet}, and hence
we have a good reason to explore this possibility from a theoretical point of view.

It is known that the pairing symmetry of high T$_c$ cuprates is described by the $d_{x^2-y^2}$ state.
The study of superfluids with $d$-wave symmetry is important because several strongly correlated
superconductors, or so-called heavy Fermion superconductors belong to unconventional
pairing states, such as $d$, $d+id$, or $f$, etc, including high T$_c$ whose pairing mechanism is still unclear.
It is also interesting because of the richer physics associated with the many
internal degrees of freedom of the relative orbital angular momentum $l>1$ of the Cooper pairs
 compared to the $l=0$ ($s$) and $l=1$ ($p$) cases.
In particular,  the spatial structure of the order parameter (OP), or the textures and vortices,
which are hallmarks of superfluidity, are intriguing in the confined ultracold atomic gases.

We note here that the OP structures of the $d$-wave superfluid of fermionic atoms and spinor condensates of
bosonic spin-2 atoms are mathematically very similar because both OP's have 5 components.
Whereas the latter has been extensively investigated already\cite{yip,ueda}, the former has not been studied so far
in connection with neutral atom gases.  As we will see soon, 
the boundary conditions due to the harmonic trap are different between the two cases in an essential way,
giving rise to novel textures and vortices in the former.
This difference arises because of the different origin of the OP degeneracy 
due to the internal degrees of freedom: The orbital angular momentum lives in real space in
the former unlike the spin degrees of freedom in the latter although both are of the same $SO(3)$ symmetry
in a  homogeneous system.

Previously, in his seminal work,  Mermin\cite{mermin} presents a general framework based on Ginzburg-Landau (GL) 
formalism to describe a $d$-wave superfluid in an infinite bulk system.  He exhaustively classifies
the ground state phase diagram into three regions; ferromagnetic (FM), polar (PO) and cyclic (CY) phases, 
depending on the coupling constants
or fourth order coefficients in the GL functional.
The cyclic phase, which we focus on in this paper, is the most intriguing one because 
fermion superfluids with $s$-wave ($l=0$) or spinless $p$-wave ($l=1$) pairings produced
by a Feshbach resonance via a magnetic sweep, or the spin-1 spinor BEC\cite{ohmi,ho} do not support this phase
which only appears above $l\geq2$.

The main motivations of this paper are to provide the fundamental physical properties of 
a $d$-wave superfluid confined by  a harmonic potential in two-dimensions (2D)
in order to help to identify the $d$-wave nature experimentally and to present a non-Abelian 1/3-vortex 
which is the energetically favored state under external rotation. 
The existence of a 1/3-vortex provides a spectroscopic means to characterize 
$d$-wave superfluidity. Mutually non-commutative 1/3-vortices themselves are
already pointed out to be allowed topologically in an $F=2$ spinor BEC by several authors\cite{1/3}.
However, there is no serious calculation to examine its stability from energetic point of view
neither in spinor BEC nor in $d$-wave fermion superfluids, which is one of our main purposes in this paper.

The non-Abelian 1/3-vortices might turn out to be useful for quantum computation 
because a spatial arrangement of mutually non-commutative 1/3-vortices can be used for storing information,
namely they can be used as a topologically protected qubit.
This is somewhat similar to the idea which utilizes the Majorana particles bound to the vortex core in chiral
$p+ip$ superconductors\cite{sarma}.
Note that the 1/3-vortex discussed here does not support Majorana zero energy particles.

Our arguments are based on GL theory which is general enough to describe a $d$-wave superfluid in terms of
the free energy expanded up to fourth order in the OP. GL theory is valid near the transition temperature $T_c$, 
however, its range of applicability is known to be wider empirically. The OP $\Delta$ for a $d$-wave superfluid is 
spanned by the spherical harmonics $Y_{lm}(\hat{\bf k})$ ($\hat{\bf k}$ is a unit vector on the Fermi sphere)

\begin{eqnarray}
\Delta({\bf r})=A_m({\bf r}) Y_{lm}(\hat{\bf k})=\hat{\bf k}_iB_{ij}\hat{\bf k}_j
\end{eqnarray}

\noindent
with $l=2$, $m=-2,\cdots, 2$ and $i,j=1,2,3$. The repeated indices are summed over.
The coefficients $A_m({\bf r})$ are a complex valued functions of ${\bf r}=(x,y)$.
To emphasize the essential points, we consider a two-dimensional system assuming 
that the obtained objects extend uniformly to the third dimension.
Alternatively we may use the symmetric traceless 3$\times$3 matrix $B_{ij}$.
The $B$-matrix which has five independent elements  is convenient in
 constructing the GL functional for the OP with $SO(3)$
symmetry in addition to $U(1)$ gauge symmetry. In the following we use these two notations  interchangeably.
The bulk GL free energy functional $f_{bulk}$ is derived by Mermin as 

\begin{eqnarray*}
f_{bulk}=\alpha Tr B^{\ast}B+\beta_1|TrB^2|^2
\end{eqnarray*}
\begin{eqnarray}
+\beta_2(TrB^{\ast}B)^2+\beta_3Tr(B^2B^{\ast 2}),
\end{eqnarray}

\noindent
where $\alpha(T)=\alpha_0(T-T_c)$ and $T_c$ is the transition temperature.
There are three independent fourth order terms $\beta_1$, $\beta_2$ and $\beta_3$.
The trace operation for a matrix is denoted by $Tr$.
This bulk energy $f_{bulk}$ is recast into

\begin{eqnarray*}
f_{bulk}=\alpha_0(T-T_c)\Sigma_i |A_i|^2+{15\over 2}\{\beta_2+{1\over 3}\beta_3
\end{eqnarray*}
\begin{eqnarray}
+(\beta_1+{1\over 6}\beta_3)|\Theta|^2-{1\over 12}\beta_3|\vec f|^2\}\Sigma_i|A_i|^4
\end{eqnarray}

\noindent
where the orbital singlet pairing amplitude $\Theta=(-1)^iA_iA_{-i}/|A_l|^2$ and 
the orbital momentum ${\vec f}=A_i^{\ast}{\vec F}_{ij}A_j/|A_l|^2$
with ${\vec F}$ being the 5$\times$5 spin matrix\cite{isoshima}.
The three phases are characterized by FM ($\langle\Theta\rangle=0,\langle{\vec f}\rangle\neq 0$),
CY ($\langle\Theta\rangle=0,\langle{\vec f}\rangle=0$), PO ($\langle\Theta\rangle\neq 0,\langle{\vec f}\rangle= 0$)
where $\langle\cdots\rangle$ is the ground state average.
It is clear from Eq.~(3) that CY is stable when $\beta_1+{1\over 6}\beta_3>0$ and $\beta_3<0$, 
while FM and PO phases occupy the other regions of the $(\beta_3, \beta_1)$ parameter space
(see Fig. 1 in Ref.\cite{mermin}).
The canonical CY phase is of the form $\Delta(\hat{\bf k})=iY_{22}(\hat{\bf k})+\sqrt2 Y_{20}(\hat{\bf k})+iY_{2-2}(\hat{\bf k})$
or in a vectorial notation: $(A_2,A_1,A_0,A_{-1},A_{-2})^T=(i,0,\sqrt2,0,i)^T$. The other cyclic states are obtained
from this by rotations.
The weak coupling estimate for the Fermi sphere ($\beta_2=2\beta_1$ and $\beta_3=0$)
predicts that the stable phase is on the boundary between the CY and FM phases.

The gradient terms are constructed by considering the possible contractions of 
$(\partial_iB_{jk})^{\ast}$ and $(\partial_lB_{mn})$; namely
there are three independent terms: (1) $(\partial_iB_{jk})^{\ast}(\partial_iB_{jk})$
(2) $(\partial_iB_{jk})^{\ast}(\partial_kB_{ij})$
and (3) $(\partial_jB_{ij})^{\ast}(\partial_kB_{ik})$.
This can be understood from the fact that the decomposition of the two
angular momenta $(L=1)\times (L=2)\rightarrow L=3,2,1$ where the former (latter) corresponds
to $\partial$ (OP).
Then, the gradient energy\cite{note1,note2} can be summed up as 

\begin{eqnarray*}
f_{grad}=K_1(\partial_iB_{jk})^{\ast}(\partial_iB_{jk})+K_2(\partial_iB_{jk})^{\ast}(\partial_kB_{ij}),
\end{eqnarray*}
\begin{eqnarray}
+K_3(\partial_jB_{ij})^{\ast}(\partial_kB_{ik})
\end{eqnarray}

\noindent
where $K_1=K$, $K_2=2K$, $K_3=2K$ with $K=2K_0/35$ in the weak coupling approximation
which  also gives $\alpha=2\alpha_0 /15$, $\beta_2=8\beta_0 /315$, 
OP amplitude $\Delta_0=\sqrt{\alpha_0 /2\beta_0}$
and the coherence length $\xi_0=\sqrt{K_0/ \alpha_0}$. (For $\alpha_0$, $\beta_0$, and $K_0$, see \cite{tsutsumi}).
We can derive the result of Eq. (4) also starting from the $A_m$ representation after simple but tedious calculations.

In addition to these bulk and gradient terms, we take into account a harmonic trap potential:

\begin{eqnarray}
f_{harmonic}= \omega^2{\bf r}^2 \Sigma_i |A_i|^2.
\end{eqnarray}

\noindent
External rotation may be included  by replacing ${\vec \partial}\rightarrow{\vec \partial}-i{2m\over \hbar}{\bf \Omega}\times{\bf r}$
where the rotation axis ${\bf \Omega}\parallel {\bf z}$. Therefore the resulting total GL functional is given by

\begin{eqnarray}
f_{total}=f_{bulk}+f_{grad}+f_{harmonic}+f_{cent}
\end{eqnarray}

\noindent
where $f_{cent}$ is the centrifugal potential\cite{tsutsumi}.
The associated mass current is derived as 

$$j_i=2{\rm Im}[K_1B^{\ast}_{jk}\nabla_iB_{jk}+K_2B^{\ast}_{jk}\nabla_kB_{ij}+K_3B^{\ast}_{ij}\nabla_kB_{jk}].$$

The following numerical calculations are done in two-dimensional plane of cartesian coordinates $(x,y)$.
The coupled GL equations for five components are solved on discretized lattice using 101$\times$101 meshes via a simple iteration.
The OP amplitudes and length are scaled by $\Delta_0$ and $\xi_0$ respectively. The trap frequency $\omega$ is
normalized by ${7\over 3}{h/2m\over 2\pi \xi_0^2}$ with $m$ the atomic mass. 
We fix $\omega=1/25$ in these units throughout this paper.

Before discussing the CY and 1/3-vortex, we briefly touch on the weak coupling case,
which indicates the stable phase on the boundary between FM and CY for an  infinite system as mentioned,
in order to understand the importance of the boundary condition through the gradient terms. 
We have done numerical calculations for a trapped system by solving the GL equations for 
five components. It turns out that FM is energetically favorable over CY. This is because of the boundary condition,
which strongly constraints the possible phase. In fact, FM is more flexible than CY in the sense
that the $\vec f$-vectors can arrange to stabilize the state.
As seen from Fig.~1 where the $\vec f$-vector configuration is shown for  FM, 
the $\vec f$-vectors tend to align along the circumference. This arrangement is energetically advantageous 
because the four point nodes situated along the $\vec f$ direction tend to lie outside the system, 
leading to maximal gain in the condensation energy.
This is analogous to that in $p$-wave superfluids\cite{tsutsumi}.
This flexible feature is absent in CY because ${\vec f}=0$.
Thus under confinement the degeneracy between FM and CY is lifted, the latter being never stabilized over FM
in the weak-coupling limit.

\begin{figure}
\includegraphics[width=4cm]{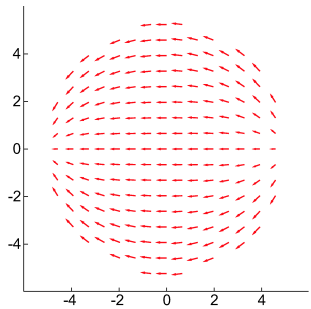}
\caption{(Color online) 
Stable ferromagnetic texture (FM) in the weak coupling case at rest.
The $f_x$-$f_y$ vectorial pattern is shown. The length is scaled
by $\xi_0$. 
}
\end{figure}

Let us now come to our main part. 
We take the $\beta_i$ values appropriate for CY,
for example,  $\beta_1/\beta_2=1$, $\beta_3/\beta_2=-1$.
The following results do not depend on these values and are generic in the
cyclic region.
By solving the GL equations, we obtain the stable CY in the presence of a trap. Among various
CY forms derived from the canonical CY we stabilize the particular CY phase, called CY-z
described by $(1,0,0,\sqrt2,0)^T$, which is obtained by 
$\exp(i\cos^{-1}({1\over \sqrt3}){\vec F}_y)\exp(-i{\pi\over 4}{\vec F}_z)(i,0,\sqrt2,0,i)^T$.
This CY-z  will be seen to support the 1/3-vortex later.
As seen from the main panel of Fig.~2 where the cross sections of each component are displayed,
CY-z is dominant in the central region because $\langle{\bf f}\rangle=0$ and simultaneously $\langle{\Theta}\rangle=0$ 
as shown in Figs.~2(a) and (b). The surface region consists of FM or PO phases, which are intermingled.
Note that FM is advantageous in the surrounding region because as mentioned 
the system can gain in the condensation energy by tuning the direction of $\vec f$. 
This OP texture differs completely  from the spinor F=2 BEC where the CY-z extends all the way to
the surface of the cloud\cite{jukka} because of different boundary conditions.
 We emphasize that wherever OP  varies spatially, the gradient coupling terms $f_{grad}$ 
in Eq.~(4) inevitably induce other components.

The gap structure of CY consists of 8 point nodes;
In canonical CY $(i,0,\sqrt2,0,i)^T$ the point nodes are situated at the 8 directions $(\pm1,\pm1,\pm1)$
given by 8 corners of the inscribed cube inside the Fermi sphere. In CY-z the
inscribed cube is rotated so that the (1,-1,1) direction lies now parallel to the $z$ axis.
Thus this CY-z is three-fold symmetric with respect to combined gauge transformation and
rotations about the  $z$ axis.
This is the origin for the possible existence of a 1/3-vortex in CY-z as discussed below.
The mass current given above flows spontaneously circularly around the center
of the trap and gradually decreases farther away from the origin.

\begin{figure}
\includegraphics[width=7cm]{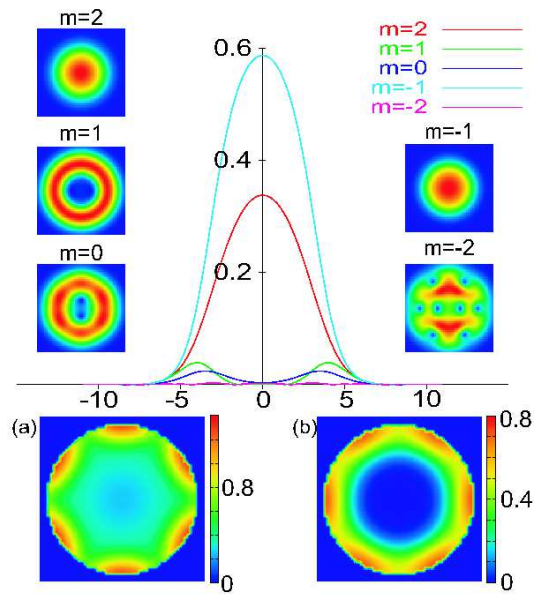}
\caption{(Color online) 
Stable texture of the cyclic phase (CY-z) at rest.
Cross section is shown along the
horizontal direction together with their contour maps
 of the five components as seen from above.
$|\vec f|$ (a) and $|\Theta|$ (b) are displayed. 
The field of view is $12\xi_0\times12\xi_0$.
}
\end{figure}

By rotating CY-z we can create the 1/3-vortex whose main structure is described by
$(e^{i\theta},0,0, \sqrt2,0)^T$ in the central region as seen from Fig.~3
where the cross section for each component is displayed.
Note that at the center the $m=+2$ component vanishes whereas the $m=-1$ component is non-zero.
Simultaneously the other components are induced around the surface region.
Combination of  the three-fold symmetry around the $z$-axis for CY-z
with an additional gauge transformation leads to the 1/3-vortex form.
Namely, CY-z $Y_{2,2}+\sqrt2Y_{2,-1}$ transforms into
$Y_{2,2}e^{2i\phi}+\sqrt2Y_{2,-1}e^{-i\phi}$ under a rotation of an angle $\phi$
around the $z$-axis, which can be rewritten as 
$e^{-i\phi}(Y_{2,2}e^{3i\phi}+\sqrt2Y_{2,-1})=e^{-i\theta/3}(Y_{2,2}e^{i\theta}+\sqrt2Y_{2,-1})$
after identifying $\phi=\theta/3$. This is apparently of the 1/3-vortex form $(e^{i\theta},0,0,\sqrt2,0)^T$.
The vortex center is dominated by the $A_{-1}$ component
where the  $A_{2}$ component vanishes due to the phase singularity
at the core as seen from Fig.~3. Thus the core of the 1/3-vortex is ferromagnetic.
The nodal structure of the ferromagnetic core region is characterized by a line node on
the equator of the Fermi sphere in addition to point nodes at both poles
because the corresponding basis function is $Y_{2,-1}\propto k_z(k_x-ik_y)$.

We also calculated the mass current for the 1/3-vortex and the line integral of the
velocity field $\oint v_{\theta}d\theta$ along various closed paths around the
center. We find that the line integral yields approximately ${1\over 3}{h\over 2m}$ when circling near the
center, implying 1/3 quantization. However, this conclusion is not exact because the system is inhomogeneous,
 a situation different from the infinite system where the quantization is exact if
the line integral is taken along a path far away from the vortex line.

\begin{figure}
\includegraphics[width=7cm]{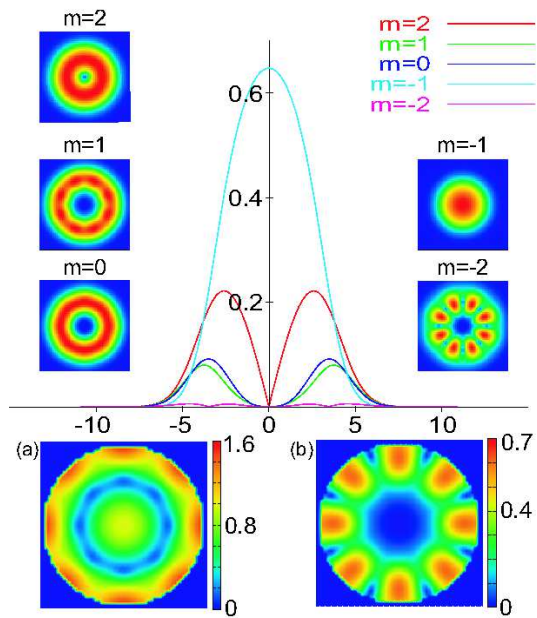}
\caption{(Color online) 
The 1/3-vortex is stable at $\Omega=0.20\omega$.
Main panel shows the cross section for five components
along the horizontal direction together with their contour maps
as seen from above. $|\vec f|$ and $|\Theta|$ are displayed in (a) and 
(b), respectively. The field of view is $12\xi_0\times12\xi_0$.
}
\end{figure}

\begin{figure}
\includegraphics[width=7cm]{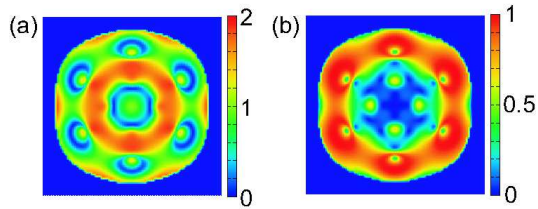}
\caption{(Color online) 
Stable texture at high rotation at $\Omega=0.28\omega$.
$|\vec f|$ (a) and $|\Theta|$ (b) are displayed
where the central region is a 1/3-vortex surrounded by
different kinds of vortices. The field of view is $20\xi_0\times20\xi_0$.
}
\end{figure}

It is remarkable that this 1/3-vortex is the most stable vortex among various possible ones
which we have solved in order to compare their energies.
The critical rotation frequecy is found to be
$\Omega_{cr}=0.24\omega$ at $T=0.5T_c$ beyond which a single 1/3-vortex becomes energetically stable
compared to the CY-z.
Upon increasing $\Omega$ further, 
different kinds of vortices emerge from the surface as shown in Fig.~4.
They are similar to the half-quantum vortex (HQV) seen
in $p$-wave  superfluids\cite{tsutsumi}.
At the center the 1/3-vortex exists surrounded by 8 HQV like objects.
 In order to create non-commutative 1/3-vortices in a system and to observe non-Abelian
 braiding statistics among them,
we need a special devise, for example, after creating a 1/3-vortex 
we should change the rotation axis from the z-axis to a different one.

The initial preparation of CY-z itself is not difficult where the appropriate atom population ratio (1:2) in
$A_2$ and $A_{-1}$ is needed. Since these two components are different orbital angular momentum states,
the population in the components can be adjusted using Raman transitions, similarly as 
in F=2 spinor BEC\cite{bigelow} or by using  RF transition in the presence of Zeeman splitting of the 
5 components under an applied 
field as was done by Hirano group\cite{hirano}.
Furthermore, once CY-z is prepared, it is possible to imprint the unit phase winding in the 
$A_2$ component using a circular polarized Raman transition\cite{bigelow}.
It may be difficult to use the optical spoon method to create the vortex, which was utilized to produce
vortices in scalar BECs\cite{madison}.

We thank T. Ohmi, M. Ozaki, T. Mizushima, M. Kobayashi, T. Kawakami, T. Hirano and R. Hulet
for useful discussions.

During preparing the manuscript, we became aware of related preprints\cite{demler,wu}.
The former (latter) treats $d$ ($f$)-wave pairing in atomic gases from different points  of view than here.


\begin{thebibliography}{99}
\bibitem{note2}
We also note that in the $F=2$ spinor BEC the gradient term is uniquely given by
$(\partial_iB_{jk})^{\ast}(\partial_iB_{jk})$ because the partial derivative operators
act on real space while the OP lives in spin space.
\end{thebibliography}
\end{document}